\documentclass[]{aastex63}

\revised{25 July, 2022}
\accepted{27 July, 2022}
\submitjournal{AJ}

\shortauthors{Ye et al.}

\begin{document}

\title{Revisit of open clusters UPK 39, UPK 41 and PHOC 39 : a new binary open cluster found}

\correspondingauthor{Jingkun Zhao}
\email{zjk@nao.cas.cn}

\author[0000-0002-5805-8112]{Xianhao Ye}
\affiliation{CAS Key Laboratory of Optical Astronomy, National Astronomical Observatories, Chinese Academy of Sciences, Beijing 100101, China}
\affiliation{School of Astronomy and Space Science, University of Chinese Academy of Sciences, Beijing 100049, China}

\author[0000-0003-2868-8276]{Jingkun Zhao}
\affiliation{CAS Key Laboratory of Optical Astronomy, National Astronomical Observatories, Chinese Academy of Sciences, Beijing 100101, China}

\author[0000-0002-8541-9921]{Terry D. Oswalt}
\affiliation{Embry-Riddle Aeronautical University, 1 Aerospace Boulevard, Daytona Beach, FL 32114, USA; oswaltt1@erau.edu}

\author[0000-0001-7609-1947]{Yong Yang}
\affiliation{CAS Key Laboratory of Optical Astronomy, National Astronomical Observatories, Chinese Academy of Sciences, Beijing 100101, China}
\affiliation{School of Astronomy and Space Science, University of Chinese Academy of Sciences, Beijing 100049, China}

\author[0000-0002-8980-945X]{Gang Zhao}
\affiliation{CAS Key Laboratory of Optical Astronomy, National Astronomical Observatories, Chinese Academy of Sciences, Beijing 100101, China}
\affiliation{School of Astronomy and Space Science, University of Chinese Academy of Sciences, Beijing 100049, China}

\begin{abstract}

We investigate the three open clusters near Aquila Rift cloud, named as UPK 39 (\texttt{c1} hereafter), UPK 41 (\texttt{c2} hereafter) in Sim et al. (2019) and PHOC 39 (\texttt{c3} hereafter) in Hunt \& Reffert (2021), respectively. Using photometric passpands, reddening, and extinction from Gaia DR3, we construct the color-absolute-magnitude diagram (CAMD). Using isochrone fits their ages are estimated as $6.3\pm0.9$, $8.1\pm1.4$ and $21.8\pm2.2$ Myr, respectively. Their proper motions and radial velocities, estimated using data from Gaia and LAMOST are very similar. From their orbits, relative distances among them at different times, kinematics, ages, and metallicities, we conclude that \texttt{c1} and \texttt{c2} are primordial binary open cluster, which are likely to have been formed at the same time, and \texttt{c3} may capture \texttt{c1}, \texttt{c2} in the future.

\end{abstract}


\section{Introduction} \label{sec:intro}

Open clusters (OCs), gravitationally bound stars originally formed from giant molecular clouds (GMCs; \citealt{Lada03}), are building blocks of the Milky Way. Catalogs of OCs have been compiled for over a century \citep{Dreyer1888}. High quality astrometric and photometric data from Gaia \citep{Gaia16,gaia_summary_dr2,gaia_summary_edr3}, combined with new highly efficient tools such as \texttt{DBSCAN} \citep{dbscan}, \texttt{HDBSCAN} \citep{Campello13,McInnes17}, \texttt{UPMASK} (Unsupervised Photometric Membership Assignment in Stellar Clusters) \citep{upmask}, have dramatically increased the number of OCs \citep{CG18,CG19,Castro18,Castro19,Castro20,Castro22,LP19,Sim19,Bica19,Ferreira19,Ferreira21,He21,He22,Hunt21,Qin21}. A complete census of OCs within the solar neighborhood will provide a sound basis to investigate a number of scientific questions. OCs could support us investigate other scientific program as well, such as the metallicity gradient of the Galaxy and radial migration \citep{Netopil22,Zhang21,Chen20}, spiral arms \citep{Castro21b}, and moving groups \citep{Zhao09,Liang17,Yang21}. With more and more identified OCs, delicate analysis for them is needed to provide further understanding of the Galaxy.

\begin{figure*}[htbp!]
  \centering
  \includegraphics[width=1.00\textwidth]{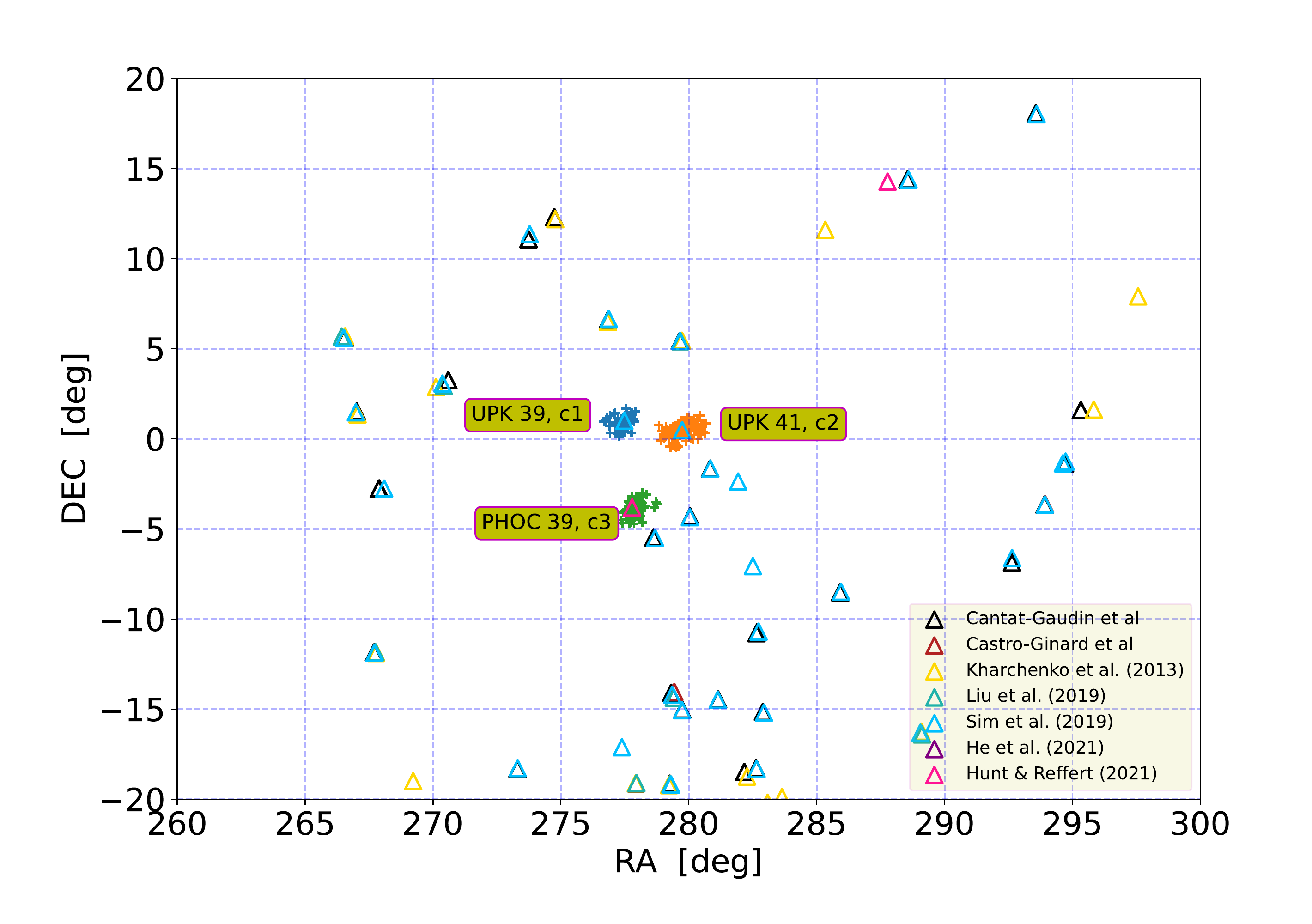}
	\caption{Pluses (+) represent stars in the three OCs \texttt{c1}(blue), \texttt{c2}(orange) and \texttt{c3}(green) of our study, located in 5D phase space. The corresponding color is adopted
  through this paper. Other known OCs from literature in this part of the sky are represented by colored triangles. \label{fig:extraction}}
\end{figure*}

In the region around the Aquila Rift cloud, we identified three OCs with similar proper motions (\texttt{PMs}). Two of them, \texttt{c1}, \texttt{c2}, were first found by \cite{Sim19} and the other, \texttt{c3} \cite{Hunt21}. \cite{Sim19} identified the centers of \texttt{c1} and \texttt{c2}, respectively, as $\left( \mu_{\alpha}^{*},\mu_{\delta} \right) = \left( 3.29\pm0.39,-8.65\pm0.41 \right)$ $\mathrm{mas} \cdot \mathrm{y}^{-1}$ and $\left( 2.51\pm0.22,-8.13\pm0.23 \right)$ $\mathrm{mas} \cdot \mathrm{y}^{-1}$. \cite{Hunt21} located the center of \texttt{c3} as, $\left( \mu_{\alpha}^{*},\mu_{\delta} \right) = \left( 1.89\pm0.05,-8.75\pm0.05 \right)$ $\mathrm{mas} \cdot \mathrm{y}^{-1}$. \cite{Sim19} estimated the ages of \texttt{c1} and \texttt{c2} to be about $2.8$ Myr and $7.1$ Myr, respectively. As these three OCs are close to each other, young, and have similar \texttt{PMs}, we are interested in their detailed dynamic properties, such as the possibility they are gravitationally bound or interacting.

A certain percentage ($\sim 12\%$) of OCs comprise binary or multiple systems in solar neighborhood \citep{dela09}, either primordial or captured. At the present rapid pace of OC discovery in general, one can expect new binary and multiple OCs to be identified.

In this paper, we estimate the ages of \texttt{c1}, \texttt{c2} and \texttt{c3} from isochrone fitting. We also estimate their three-dimensional space velocities. From these data and metallicity estimates we examine the possibility that they are or were physically associated.

The structure of this paper is as follows : Sec. \ref{sec:data} describes the data extracted from Gaia EDR3; In Sec. \ref{sec:methods} we detail the procedures of member selections of the three clusters and isochrone fitting using Gaia DR3 passpands; Their ages, kinematic properties, relative distances, and metallicities are presented in Sec. \ref{sec:results}; Our conclusions are given in Sec. \ref{sec:conclusion}; This paper is summarized in Sec. \ref{sec:summary}.

\section{Data} \label{sec:data}

The Gaia EDR3 astrometric parameters \citep{gaia_summary_edr3,Lindegren_astrometric} $\left( \alpha,\delta,\varpi,\mu_{\alpha}^{*},\mu_{\delta} \right)$ are used to identify member candidates of the three clusters. We first select sources within a $40^{\circ} \times 40^{\circ}$ zone on sky, and restrict the distance via the parallaxes $\varpi$. The relative errors of $\varpi$ and \texttt{PMs} are restricted to $10\%$ to ensure the qualities of data. Furthermore, we apply the renormalized unit weight error (RUWE; \citealt{Lindegren18tech,Lindegren_astrometric}) to refine our selections. The following constraints were used to define our primary sample:

\begin{itemize}
  \item 260 $<$ $\alpha$ $<$ 300, -20 $<$ $\delta$ $<$ 20, 1/0.700 $<$ $\varpi$ $<$ 1/0.250;
  \item \texttt{parallax\underline{ }over\underline{ }error} $>$ 10;
  \item $\sigma_{\mu_{\alpha}^{*}}/|\mu_{\alpha}^{*}| < 0.1$;
  \item $\sigma_{\mu_{\delta}}/|\mu_{\delta}| < 0.1$;
  \item RUWE $<$ 1.4.
\end{itemize}

Using the above criteria, 965,430 sources which contain all three Gaia passbands $\left( G,G_{\mathrm{BP}},G_{\mathrm{RP}} \right)$ are retained as our primary sample. However, only a small fraction of this sample have radial velocities \texttt{RVs} in Gaia DR3 \citep{Katz22} and the number will drop dramatically if we restrict the relative error of \texttt{RV} too strictly. It is more reasonable to search for clusters members in 5D phase space $\left( x, y, z, \kappa \cdot \mu_{\alpha}^{*} \cdot d, \kappa \cdot \mu_{\delta} \cdot d \right)$ (3D Cartesian coordinates and 2D tangential velocity), where $\kappa=4.74047$ and $d$ is distance. The photogeometric distance from \cite{Bailer-Jones21} is applied in the calculation and the inverse of parallax $1/\varpi$ is used for the distance for those without the photometric distance from \cite{Bailer-Jones21}. In addition, we correct $\varpi$ with the code\footnote{\url{https://gitlab.com/icc-ub/public/gaiadr3_zeropoint}} from \cite{Lindegren-zp} to calculate the parallax zero point. \texttt{Galpy} \citep{Bovy15} is used to calculate the Galactic Cartesian coordinates and velocities. We set the radial distance and height of Sun in Galactocentric frame at $R_{\odot}=8.3 \quad \mathrm{kpc}$ \citep{rsun}, $Z_{\odot}=0.027 \quad \mathrm{kpc}$ \citep{zsun}, and its  velocity as $\left( U,V,W \right)_{\odot} = \left( 11.1, 12.24,7.25 \right) \quad \mathrm{km \cdot s^{-1}}$ \citep{uvwsun} relative to the Local Standard of Rest (LSR), where $V_{\mathrm{LSR}} = 240.0 \quad \mathrm{km \cdot s^{-1}}$ according to \cite{LSR}.

\section{Methods} \label{sec:methods}

\subsection{Member Selections in 5D phase space} \label{member selections}

\par There are a number of known OCs in the region covering our primary sample. We collected catalogs from \cite{CG18,CG19,CG20a,CG20b,Castro18,Castro19,Castro20,Kharchenko13,LP19,Sim19,He21,Hunt21} and show them in Fig. \ref{fig:extraction} as triangles of different colors. Note we also use parallaxes gleaned from the literature to restrict our search to OCs in the same distance range as our sample. We use the same definition as in our previous papers \citep{Ye21a,Ye21b} to calculate the number of neighbors of each star in 5D phase space. Stars within a radius of $15$ pc in $\left( x,y,z \right)$ and a radius of $1.1$ $\mathrm{km \cdot s^{-1}}$ in tangential velocity are defined as the neighbors of a given source. Applying a lower limit of $\mu + 3\sigma$ (mean value plus three times standard deviation of neighbors) retains $3,432$ stars, which includes almost all the clusters in the literature. We then test both \texttt{HDBSCAN}\citep{Campello13,McInnes17} and \texttt{DBSCAN}\citep{dbscan,sklearn12} to search for member candidates of clusters among those $3,432$ stars. Eventually, we focus on the clusters in the region of $275^{\circ} < \alpha < 285^{\circ}$, $-5^{\circ} < \delta < 4^{\circ}$. Using $\epsilon=0.5,\texttt{minPts}=30$ via \texttt{DBSCAN} in $\left( \alpha,\delta \right)$, this procedure yields the three OCs marked as colored crosses in Fig. \ref{fig:extraction}. The blue and orange crosses correspond to \texttt{c1} and \texttt{c2}. The green cluster corresponds to \texttt{c3}.

\subsection{Member refinement with Radial Velocity} \label{rv}

\begin{figure}[htbp!]
  \centering
	\includegraphics[width=0.45\textwidth]{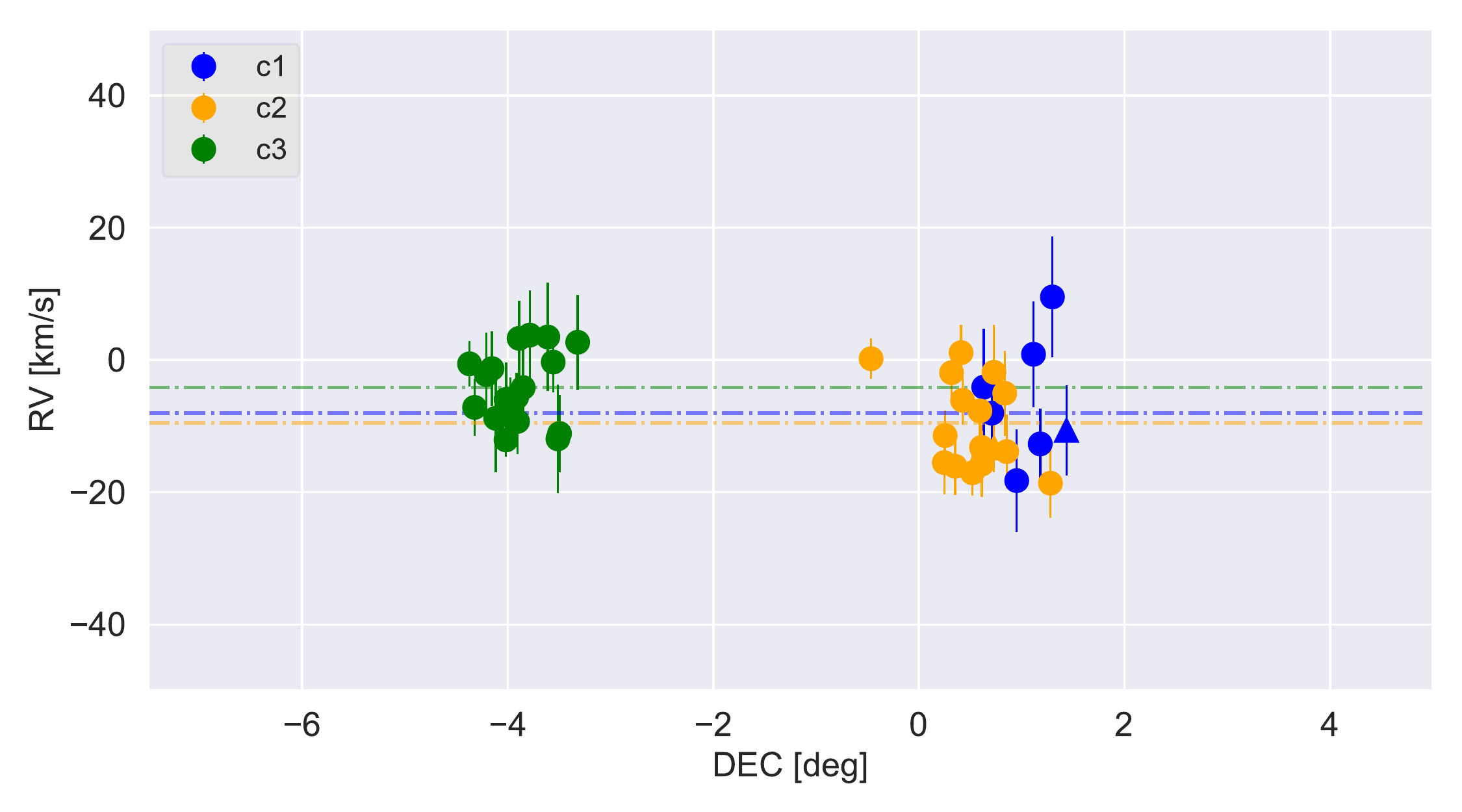}
	\caption{Dec vs. \texttt{RV} for the clusters candidates with Gaia DR3 and LAMOST DR8 \texttt{RVs}. For Gaia \texttt{RVs}, only $\sigma_{\mathrm{RV}}$ less than the mean $\sigma_{\mathrm{RV}}$ in each cluster were used. Data from LAMOST are marked as triangles. The horizontal lines indicate the final estimated \texttt{RVs} for the clusters which are the mean values. \label{fig:rv}}
\end{figure}

\begin{figure*}[t!]
  \centering
	\includegraphics[width=0.80\textwidth]{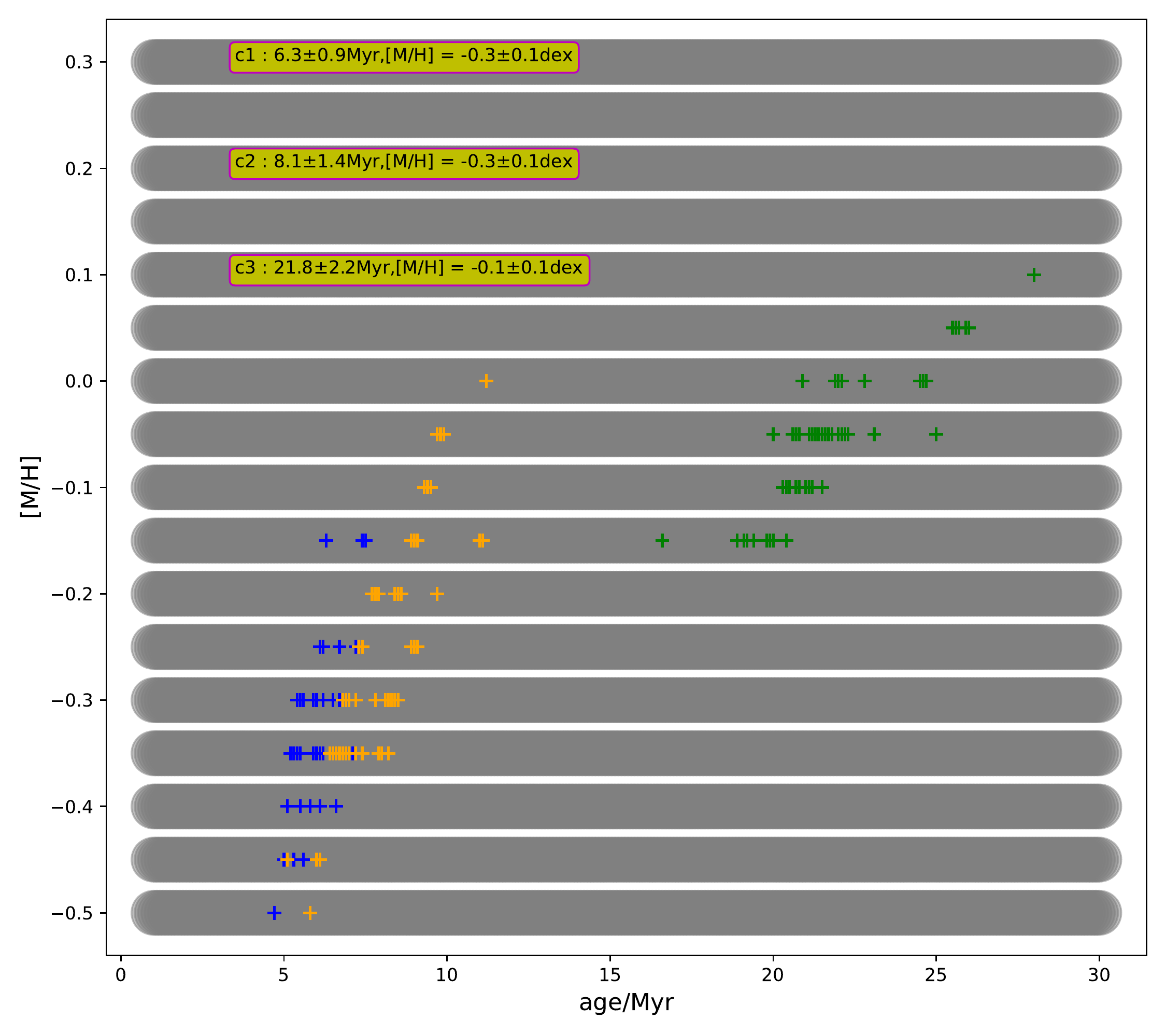}
	\caption{[M/H]] vs. age. The gray shadows show the coverages of PARSEC isochrones used in this paper. The colored crosses present the top $1\%$ of isochrones that have smallest $\bar{d^{2}}$ for the three clusters. \label{fig:cmd_age_mh}}
\end{figure*}

\par Both the literature \texttt{PMs} values and those determined for our OCs member candidates are quite similar. However, we had no prior knowledge of their radial velocities. In this paper, we found \texttt{RVs} for a few stars in Gaia DR3. The mean errors in \texttt{RVs} of the member candidates are : $10.81$, $7.26$, and $8.32$ $\mathrm{km \cdot s^{-1}}$ for \texttt{c1}, \texttt{c2}, and \texttt{c3}, respectively. We then set these values as the upper limits for $\sigma_{\mathrm{RV}}$ in our subsequent \texttt{RV} analysis. We found two stars from \texttt{c1} and \texttt{c2}, respectively, have \texttt{RVs} in LAMOST DR8 low resolution archive \citep{Liu15,Cui12,Zhao12,Zhao06}. In each cluster, some \texttt{RVs} deviate greatly from the median value, especially at $G > 14.5$ mag. Such stars are removed from our candidate list. Finally, we adopt the \texttt{RV} for each cluster as the mean value of the remaining members. Note that \texttt{c1} contains one star for which only a LAMOST \texttt{RV} is available; this star was included in the calculation of the mean \texttt{RV} for \texttt{c1}. In Fig. \ref{fig:rv}, the remaining \texttt{RVs} from Gaia DR3 are shown in Dec vs. \texttt{RV} as dots in different colors, and two \texttt{RVs} from LAMOST (in \texttt{c1} and \texttt{c2}) are marked as triangles. The adopted values are $-8.05$, $-9.53$, $-4.15$ $\mathrm{km \cdot s^{-1}}$ for \texttt{c1}, \texttt{c2} and \texttt{c3}, respectively, and are indicated with the dashed lines in different colors in Fig. \ref{fig:rv}.

\subsection{Color-absolute-magnitude Diagram and Isochrone Fitting} \label{camd}

\begin{figure*}[t!]
  \centering
	\includegraphics[width=0.90\textwidth]{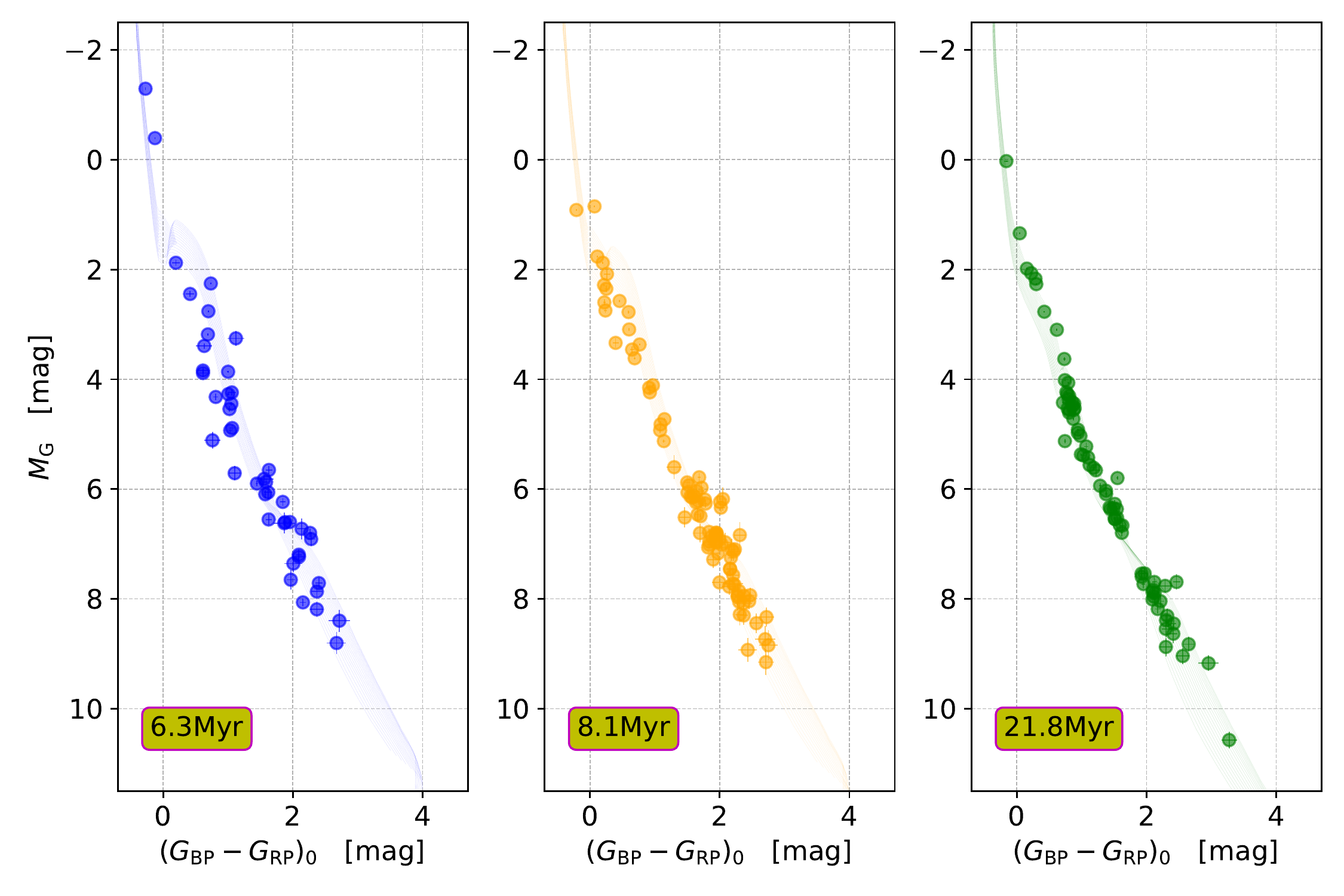}
	\caption{CAMD for \texttt{c1}, \texttt{c2} and \texttt{c3} (left to right), along with adopted ages and the corresponding isochrones with [M/H] between $-0.5$ to $0.3$ dex (interval $0.05$ dex) are shown. \label{fig:cmd_best_fits}}
\end{figure*}

\par The photometric parameters $\left( G,G_{\mathrm{BP}},G_{\mathrm{RP}} \right)$ and their errors, reddening, extinction used in this paper are from Gaia DR3, which are obtained by cross-matching the member candidates with Gaia DR3 in \texttt{TOPCAT} \citep{topcat05}. The errors in $\left( G,G_{\mathrm{BP}},G_{\mathrm{RP}} \right)$ are from CDS \footnote{\url{https://vizier.cds.unistra.fr/viz-bin/VizieR-3?-source=I/355/gaiadr3}}. For reddening \texttt{E(BP-RP)} and extinction \texttt{AG}, the uncertainties are taken as half the difference between upper and lower confidence levels, same as the uncertainties adopted in distance in this paper. There are $> 73\%$, $> 86\%$ and $> 96\%$ stars having reddening and extinction data in Gaia DR3. With the dereddened color $\left( G_{\mathrm{BP}}-G_{\mathrm{RP}} \right)_{0}$ and absolute magnitude $M_{\mathrm{G}}=G-5\lg\left(\frac{d [pc]}{10}\right)-A_{\mathrm{G}}$, where $d$ is the distance of each star, we estimated the ages of these clusters by fitting isochrones to each cluster's CAMD. Two stars in \texttt{c2} that diverge from the clearly coeval sequence were removed from the member candidate sample. The procedure outlined above yields $61$, $103$, and $78$ member stars in each cluster. In \cite{Sim19}, \texttt{c1} and \texttt{c2} were determined to have ages of about $2.8$ Myr and $7.1$ Myr, respectively, according to isochrone fitting. The main sequence of \texttt{c3} is noticeably different from the other two clusters. We use a series of PARSEC (version 1.2s) isochrones \footnote{\url{http://stev.oapd.inaf.it/cgi-bin/cmd}} \citep{Bressan12,Chen14,Chen19} with Gaia EDR3 photometric data to fit with these three clusters. The range of [M/H] of the isochrone grid is $-0.5 \sim 0.3$ dex with a interval of $0.05$ dex. The ages of isochrones range from $1$ Myr to $30$ Myr with a interval of $0.1$ Myr. The closest isochrone for a cluster is adopted by minimizing $\bar{d^2}$ in CAMD. $\bar{d^2}$ is defined as Eq. \ref{eq:cmd_d2} in \cite{LP19}. For a cluster with $n$ member candidates, $\bar{d}^2$ for a given isochrone is : 
\begin{eqnarray}\label{eq:cmd_d2}
  \bar{d^{2}} = \sum\limits_{i=1}^{n} \left| \mathbf{x}_{i} - \mathbf{x}_{i,nn} \right|^{2}/n \,,
\end{eqnarray}
where $\mathbf{x}_{i}$ gives the position of member star $i$ in CAMD, and $\mathbf{x}_{i,nn}$ represents the nearest neighbor of member $i$ in the isochrone in the same parameter space. The nearest neighbor is identified by the $k$-D tree approach in \texttt{scipy} \citep{scipy20}. The influences of the errors in $G_{\mathrm{BP}}$, $G_{\mathrm{RP}}$ and $G$ passpands have also been considered. For each star, we use the errors in $\left( G_{\mathrm{BP}}-G_{\mathrm{RP}} \right)_{0}$ and $M_{\mathrm{G}}$ to randomly generate $100$ different data points following a Gaussian distribution and calculate $\left| \mathbf{x}_{j} - \mathbf{x}_{j,nn} \right|$ for $j$ in those data points. The mean value is adopted for star $i$ as $\left| \mathbf{x}_{i} - \mathbf{x}_{i,nn} \right|$. Then $\bar{d^{2}}$ is evaluated between each cluster and each isochrone. To obtain the ages, we use the first $1\%$ of the isochrones which have the smallest $\bar{d^{2}}$. In Fig. \ref{fig:cmd_age_mh}, we present the $1\%$ isochrone age vs. [M/H]. It is apparent in this figure that \texttt{c1} and \texttt{c2} are similar both in age (\texttt{c1} : $6.3\pm0.9$ Myr, \texttt{c2} : $8.1\pm1.4$ Myr) and [M/H] (\texttt{c1} : $-0.3\pm0.1$ dex, \texttt{c2} : $-0.3\pm0.1$ dex), and \texttt{c3} is much older and richer in [M/H] than the others.

\section{Results} \label{sec:results}

\subsection{Age} \label{age}

\par The estimated ages of \texttt{c1}, \texttt{c2}, \texttt{c3} from our best-fitting isochrones are $6.3\pm0.9$, $8.1\pm1.4$ and $21.8\pm2.2$ Myr, respectively. In Fig. \ref{fig:cmd_best_fits}, cluster members and isochrones with the corresponding best ages and [M/H] between $-0.5$ to $0.3$ dex are presented.

\subsection{Galactocentric Cylindrical velocity} \label{velocity}

\begin{figure*}[htbp!]
	\includegraphics[width=0.50\textwidth]{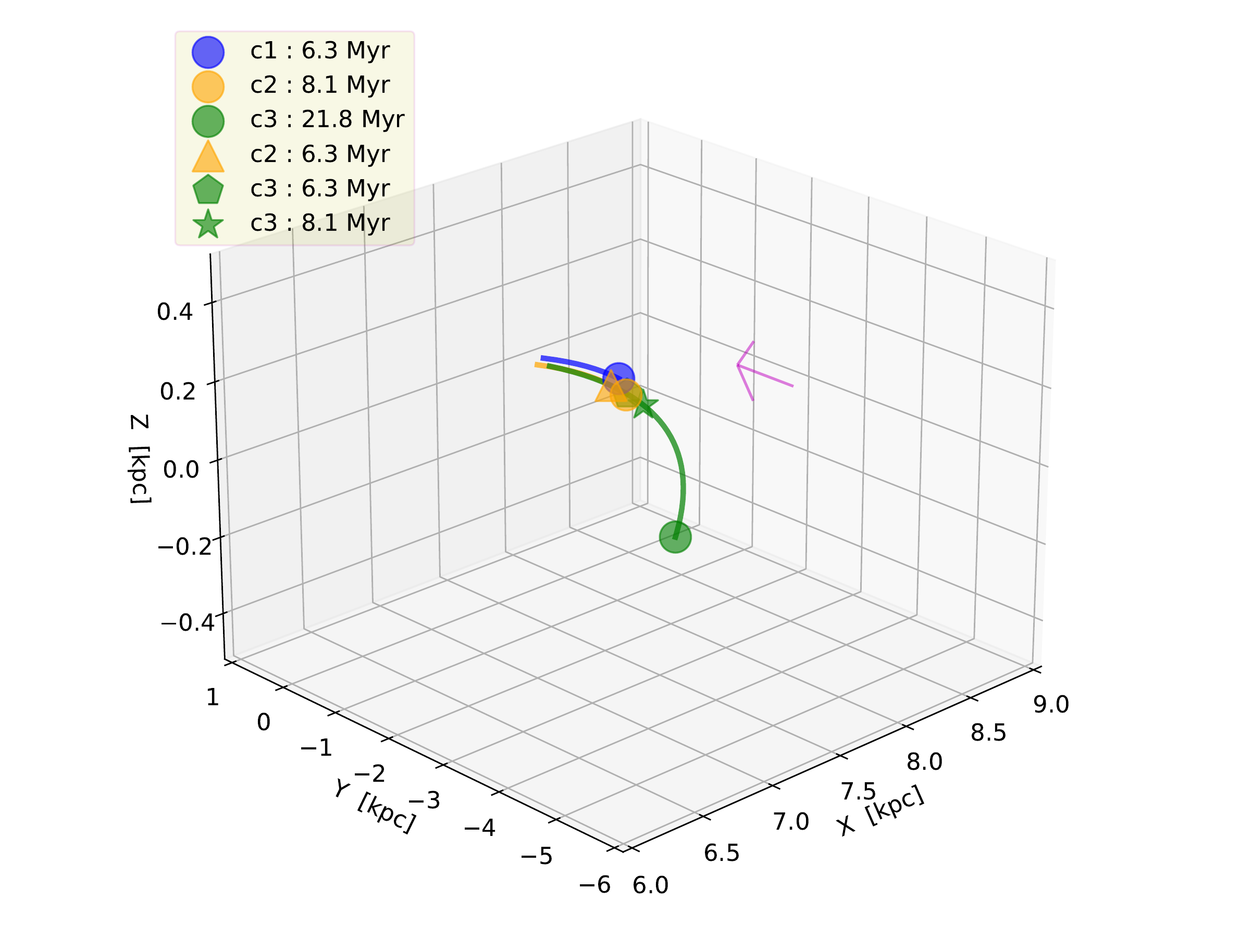}
  \includegraphics[width=0.50\textwidth]{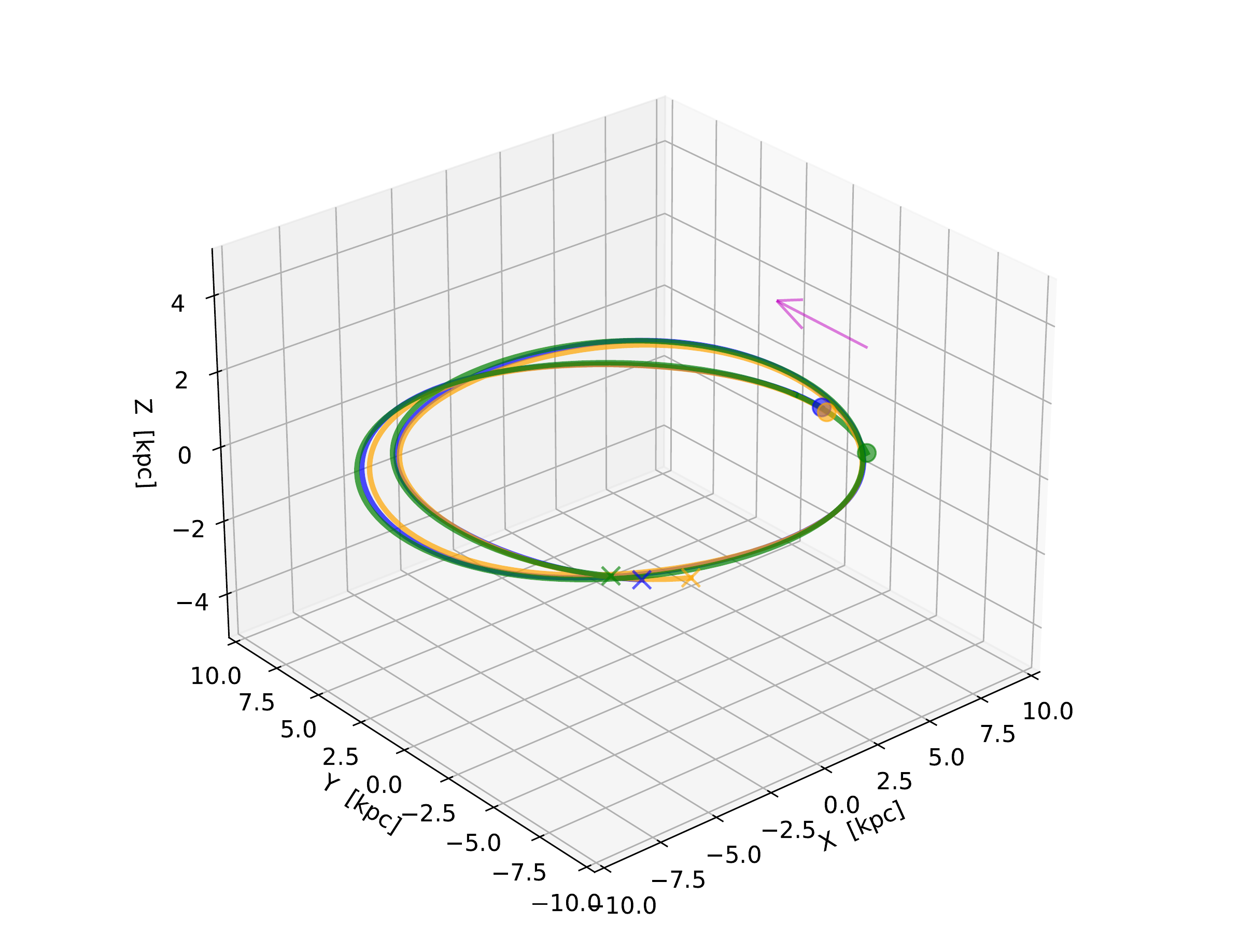}
	\caption{Left panel : Orbits of \texttt{c1}, \texttt{c2} and \texttt{c3} traced back to their birthplaces. \texttt{c3} was close to \texttt{c1} and \texttt{c2} when they were born and \texttt{c2} was near \texttt{c1} in the time of the formation of \texttt{c1}. Right panel : Circles are where these clusters were born and crosses are their locations after $400$ Myr. They did and will orbit together in the thin disk. The arrow in each panel shows the direction of motion. \label{fig:orbit}}
\end{figure*}

Averaging over our clusters members, the \texttt{PMs} of \texttt{c1}, \texttt{c2} and \texttt{c3} are $\left( \mu_{\alpha}^{*},\mu_{\delta} \right) = \left( 3.24\pm0.31,-8.59\pm0.43 \right)$ $\mathrm{mas} \cdot \mathrm{y}^{-1}$, $\left( 2.65\pm0.27,-8.35\pm0.27 \right)$ $\mathrm{mas} \cdot \mathrm{y}^{-1}$ and $\left( 2.06\pm0.20,-9.00\pm0.26 \right)$ $\mathrm{mas} \cdot \mathrm{y}^{-1}$, respectively. The \texttt{PMs} and \texttt{RVs} imply similar space velocities for these three clusters in a Galactocentric Cylindrical coordinates. We calculate that $\left( V_{\mathrm{R}},V_{\phi},V_{\mathrm{Z}} \right) = \left( -5.18,238.33,-7.17 \right)$ $\mathrm{km} \cdot \mathrm{s}^{-1}$, $\left( -3.54,236.09,-6.91 \right)$ $\mathrm{km} \cdot \mathrm{s}^{-1}$ and $\left( -8.42,239.61,-3.61 \right)$ $\mathrm{km} \cdot \mathrm{s}^{-1}$ for \texttt{c1}, \texttt{c2} and \texttt{c3}, respectively, where $V_{\mathrm{R}}>0$ represents moving away from Galactic center, $V_{\phi}$ is the direction of Galactic rotation, and $V_{\mathrm{Z}}$ points to the Galactic North Pole. 

\subsection{Orbit and Separation} \label{orbit}

\begin{figure}[b!]
  \centering
	\includegraphics[width=0.45\textwidth]{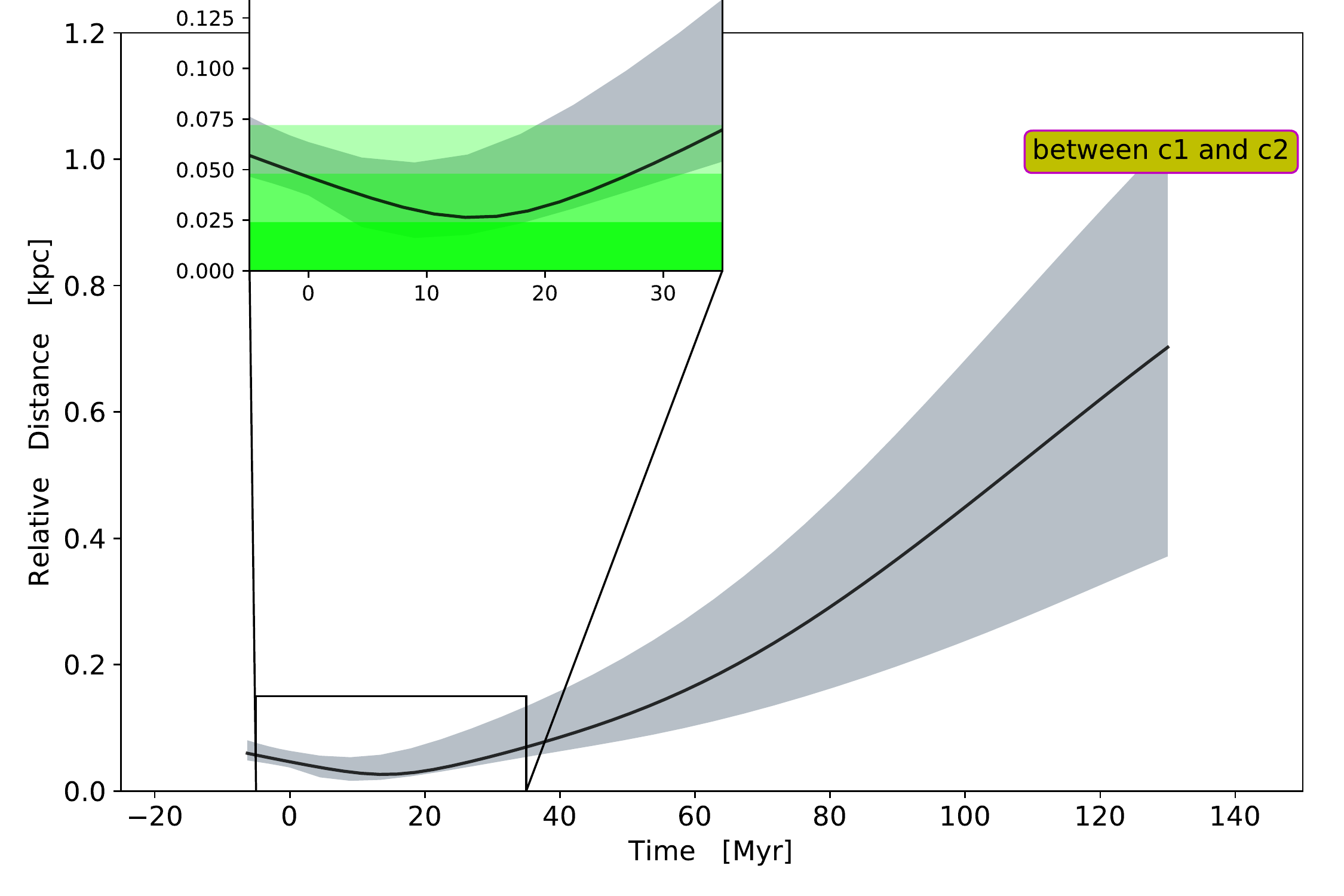}
  \includegraphics[width=0.45\textwidth]{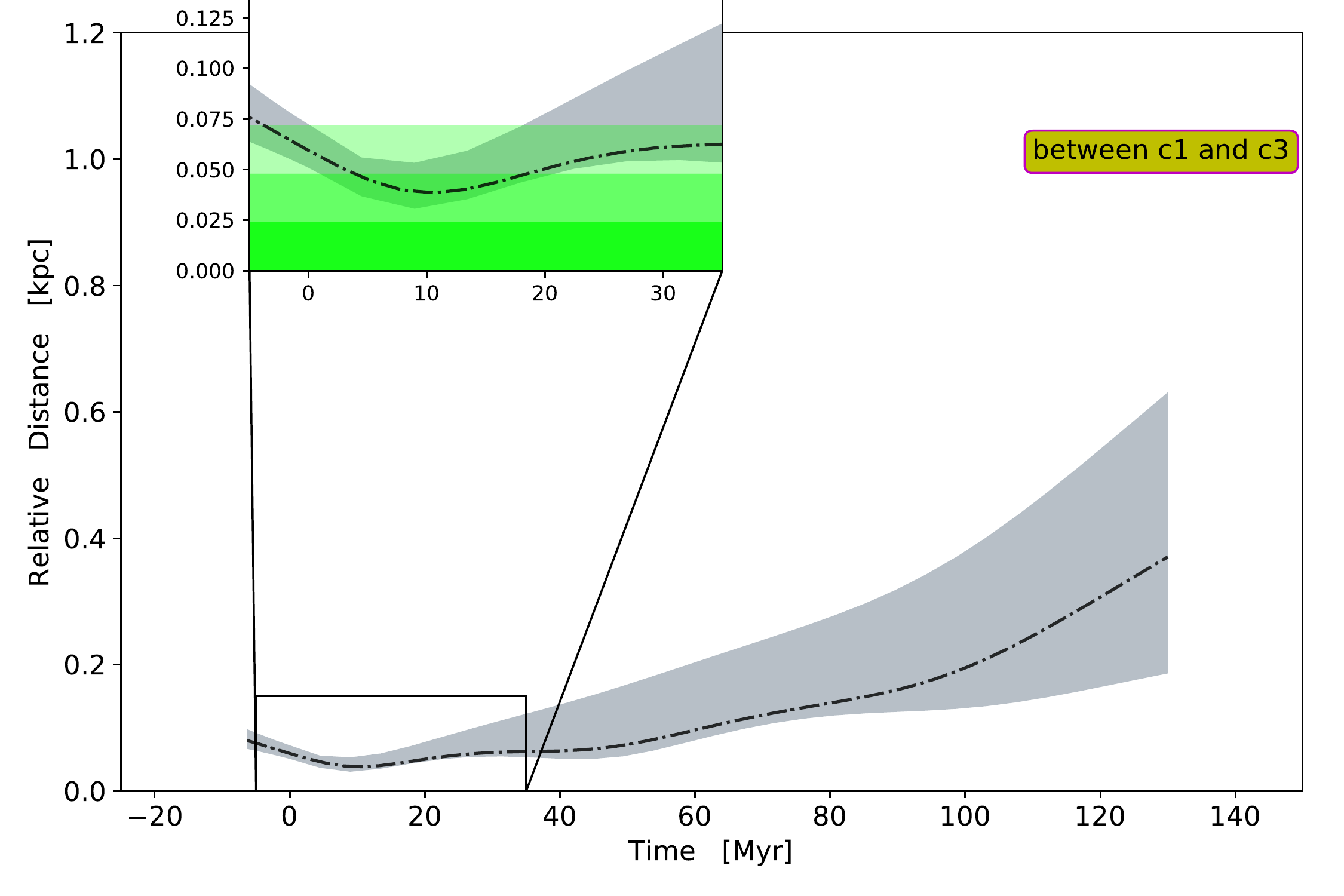}
  \includegraphics[width=0.45\textwidth]{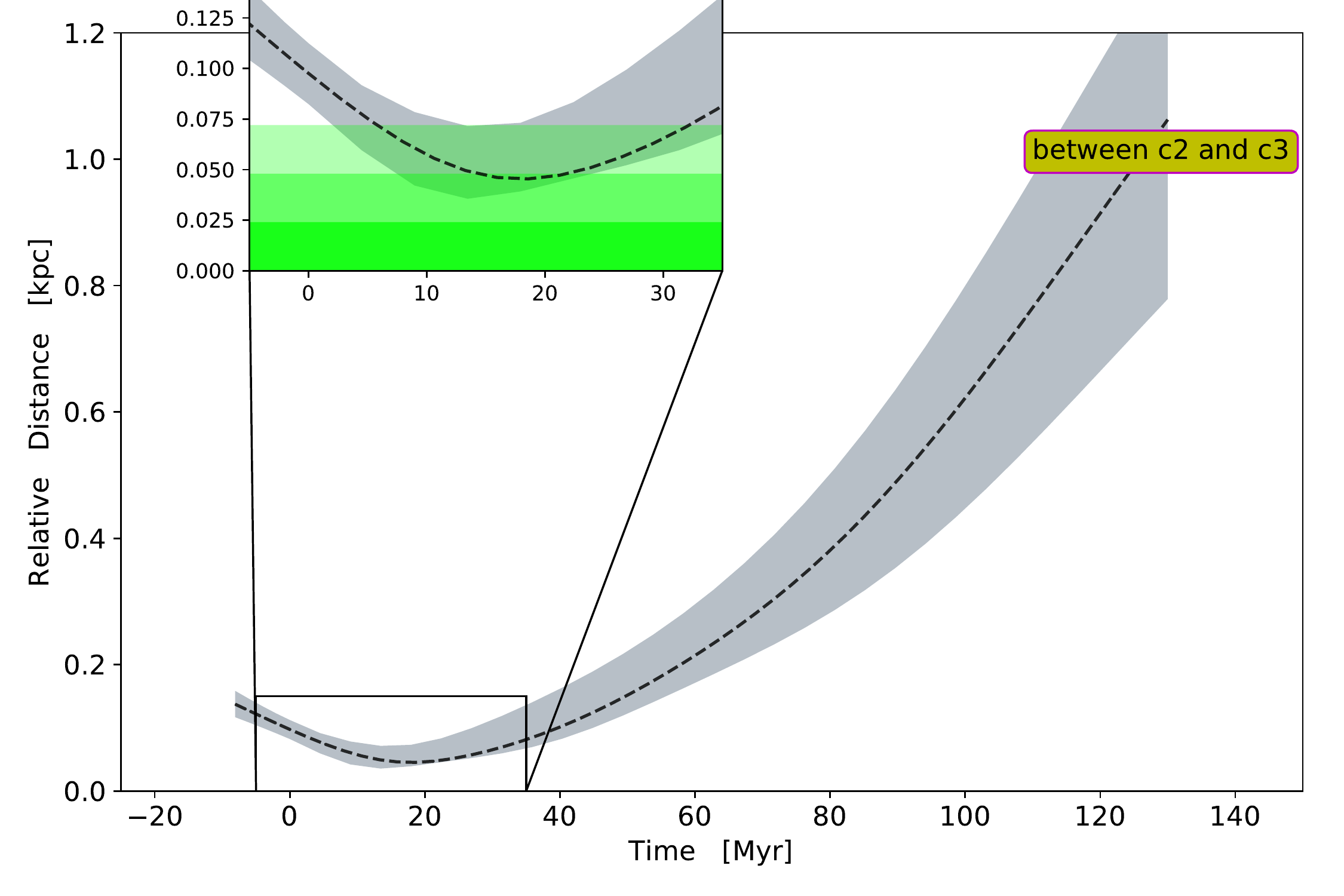}
	\caption{Relative distances among \texttt{c1}, \texttt{c2}, \texttt{c3} vs. time. The gray shaded areas present the uncertainties of relative distances and the green shaded areas denote their relative distances in multiples of $\bar{r_{\mathrm{t}}}$.  \label{fig:relative_distance}}
\end{figure}

\par As the velocities of these three OCs are very similar, we are curious about their Galactic orbits and three origins. We use \texttt{MWPotential2014} in \texttt{Galpy} to trace their birthplaces that correspond to their ages. In the left panel of Fig. \ref{fig:orbit}, we trace back \texttt{c1}, \texttt{c2}, \texttt{c3} to their birthplaces, marked as filled circles. The orange triangle shows \texttt{c2}'s position when \texttt{c1} was born. The positions of \texttt{c3} in $6.3$ Myr and $8.1$ Myr ago are presented as other green signs. The right panel of Fig. \ref{fig:orbit} presents the orbits from their birthplaces to where they will be $400$ Myr henceforth. The circles represent the origins and the crosses are their locations in the future. All the three clusters will orbit in unison in the Galactic thin disk. The minimal distances among them are on the order of dozens of parsecs. We use the standard deviations of positions and velocities of member stars in each cluster to randomly generate parameters following a Gaussian distribution. The adopted deviations in \texttt{RVs} are taken as $1.1$ $\mathrm{km} \cdot \mathrm{s}^{-1}$, the same as we used in Sec. \ref{member selections} to extract member candidates from tangential velocities. We produce three data points in each position parameters and five in each velocity parameter. We then construct $3,375$ different orbits for each cluster and calculate the relative distance among \texttt{c1}, \texttt{c2}, and \texttt{c3} at different times.

In Fig. \ref{fig:relative_distance}, the separations between two clusters are presented. The black (solid, dotted, dashed) lines in the panels are the separations calculated by the adopted positions and velocities for \texttt{c1}, \texttt{c2}, and \texttt{c3}. The gray shaded regions indicate the uncertainties in the orbits. These lower and upper bounds are $\mu-\sigma$ and $\mu+\sigma$ of the separations calculated with the generated data points. If the relative distance between two clusters is less than three times the outer radius, the mutual interaction becomes significant \citep{Innanen72}. \cite{dela09} used three times the mean tidal radius ($r_{\mathrm{t}}$) as the upper limit to select paired OCs. The adopted mean value of tidal radius is $10$ pc in \cite{dela09}. However, this value may be smaller than the actual average value. In \cite{Kharchenko13}, the mean value of $r_{\mathrm{t}}$ is $11.59$ pc as determined from an analysis of nearly $3,000$ OCs. In recent research, parts of the known OCs have coronae or haloes \citep{Meingast21}. According to \cite{Tarricq22}, the mean tidal radius is $\sim 30$ pc as inferred from an analysis of more than $300$ OCs in the solar neighborhood. This same study suggested a few OCs have $r_{\mathrm{t}} > 65$ pc. Leaving out those OCs with too large $r_{\mathrm{t}}$, the mean $r_{\mathrm{t}}$ is still $\sim 24$ pc in \cite{Tarricq22}. Some of our study's cluster members do not have reddening and extinction estimates, which are important to estimating the stellar mass in CAMD using isochrones. The member candidates of \texttt{c1}, \texttt{c2}, \texttt{c3} are clearly incomplete in $M_{\mathrm{G}}$, as seen in Fig. \ref{fig:cmd_best_fits}. Therefore, $r_{\mathrm{t}}$ calculated based on only these member candidates may be underestimated. In this work, we adopted an average tidal radius $\bar{r_{\mathrm{t}}} \sim 24$ pc. Therefore, clusters separated by less than $3 \times \bar{r_{\mathrm{t}}}$ pc may be tidally interacting. In Fig. \ref{fig:relative_distance}, the green shading with different transparencies are the regions where relative distances are in $1 \times \bar{r_{\mathrm{t}}}$, $2 \times \bar{r_{\mathrm{t}}}$ and $3 \times \bar{r_{\mathrm{t}}}$ pc.

\begin{figure*}[htbp!]
  \centering
	\includegraphics[width=0.95\textwidth]{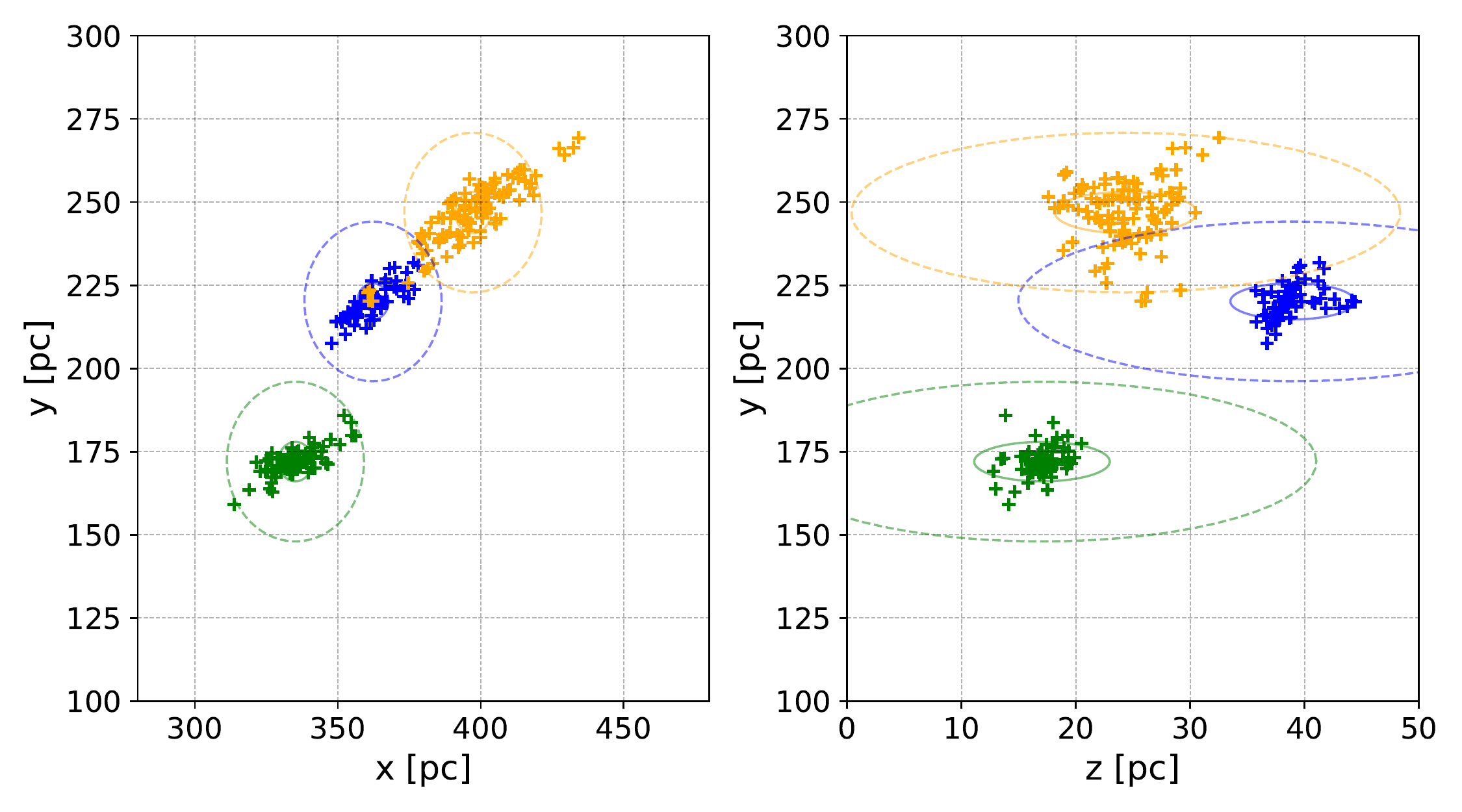}
	\caption{Distributions of member stars of \texttt{c1}, \texttt{c2} and \texttt{c3} in Cartesian coordinates $xyz$ centered at Sun. The solid circles are the tidal radii based on the total cluster mass, and the dashed circles are $\bar{r_{\mathrm{t}}} \sim 24$ pc.  \label{fig:xyz}}
\end{figure*}

The distributions of member candidates in Cartesian coordinates $xyz$ are presented in Fig. \ref{fig:xyz} to show the dimensions of the clusters and $\bar{r_{\mathrm{t}}}$ more clearly. In this figure, member stars in the $xy$ plane appear to be stretched along the line of sight, primarily due to uncertainties in distance, reddening and extinction. The solid circles are $r_{\mathrm{t}}$ calculated from the total cluster mass of \texttt{c1}, \texttt{c2} and \texttt{c3}, using Eq. \ref{rt_Pinfield} according to \cite{Pinfield98},
\begin{eqnarray}
  r_{\mathrm{t}} &=& \left[ \frac{G \mathrm{M_{total}}}{2\left( A-B \right)^2} \right]^{1/3} \,. \label{rt_Pinfield}
\end{eqnarray}
$\mathrm{M_{total}}$ in Eq. \ref{rt_Pinfield} is the accumulated stellar mass for a cluster. The other constants are : Gravitational constant $G = 4.3 \times 10^{-6}$ $\mathrm{kpc}$ $( \mathrm{km}$ $\mathrm{s}^{-1} )^{2}$ $\mathrm{M}_{\mathrm{\odot}}^{-1}$; the Oort constants $A$ = 15.3 $\pm$ 0.4 $\mathrm{kpc}^{-1}$ $\mathrm{km}$ $\mathrm{s}^{-1}$, $B$ = -11.9 $\pm$ 0.4 $\mathrm{kpc}^{-1}$ $\mathrm{km}$ $\mathrm{s}^{-1}$ from \cite{Bovy17}. The dashed circles represent the above three increments in $\bar{r_{\mathrm{t}}}$. Some members of \texttt{c1} are within the range of $\bar{r_{\mathrm{t}}}$ of \texttt{c2}, and vice versa. However, \texttt{c3} is farther apart from the other two clusters and is unlikely to be tidally influenced by them at the present time.

\begin{figure}[htbp!]
  \centering
	\includegraphics[width=0.45\textwidth]{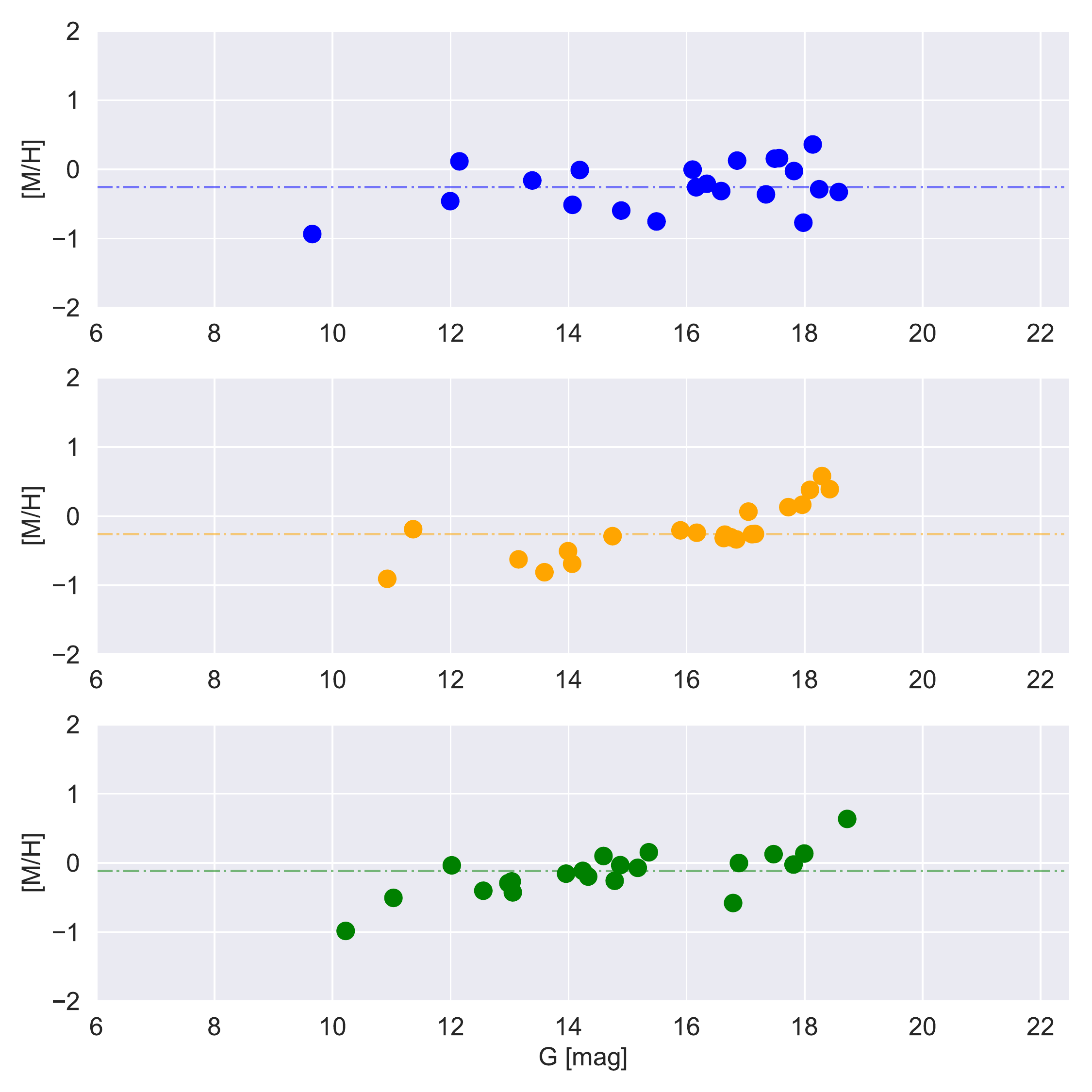}
	\caption{Distributions of [M/H] vs. $G$ for \texttt{c1}, \texttt{c2}, and \texttt{c3}. The colored lines present the median values of [M/H] for the clusters.  \label{fig:metallicity}}
\end{figure}

\subsection{Metallicity} \label{metallicity}

\par In Sec. \ref{camd}, the metallicity [M/H] was estimated from isochrone fitting : $-0.3\pm0.1$ dex for \texttt{c1}, \texttt{c2}, and $-0.1\pm0.1$ dex for \texttt{c3}. Here, we use [M/H] (\texttt{mh\underline{ }gspphot}) from Gaia DR3 to evaluate the metallicities of these clusters. We adopt the upper and lower bounds of [M/H] to represent the uncertainties of [M/H] and select $21$ member candidates of each cluster with the smallest [M/H] uncertainties, which lowers the mean uncertainties of [M/H] to $0.02$ dex for each cluster. In Fig. \ref{fig:metallicity} we present those stars in [M/H] vs. $G$, and the median values of [M/H] are $-0.26$ dex for \texttt{c1}, \texttt{c2}, and $-0.12$ dex for \texttt{c3}, indicated by the dashed lines.

\section{Conclusion}\label{sec:conclusion}

\par In this section, we discuss the probable relationships among these three OCs based on their separations, kinematics, ages and metallicities. The properties of \texttt{c1}, \texttt{c2} and \texttt{c3} are summarized in Tab. \ref{tab:c1c2c3_properties}.

\par When \texttt{c1} was formed, \texttt{c2} was very close to it. The relative distance between them is less than $2 \times \bar{r_{\mathrm{t}}}$ pc at present. The extensions of $\bar{r_{\mathrm{t}}}$ for \texttt{c1} and \texttt{c2} have some areas overlapped at present, suggesting their tidally interacting. Their $\left( V_{\mathrm{R}},V_{\phi},V_{\mathrm{Z}} \right)$, also differ by less than $3$ $\mathrm{km} \cdot \mathrm{s}^{-1}$. In addition, their ages and metallicities are almost the same. We conclude that \texttt{c1} and \texttt{c2} are a primordial binary cluster, formed simultaneously.

\par On the other hand, \texttt{c3} was formed earlier and has a higher metallicity. In addition, the current separation between \texttt{c3} and \texttt{c1} (or \texttt{c2}) and their relative space velocities suggest that there was very little chance that \texttt{c3} was closer to \texttt{c1} (or \texttt{c2}) than $\sim 3 \times \bar{r_{\mathrm{t}}}$ since \texttt{c3} was formed. Also, most massive member candidates in \texttt{c3} is no more than about $4$ $\mathrm{M_{\odot}}$. Therefore, it is unlikely that \texttt{c3} triggered the formation of the other two clusters. We consider \texttt{c3} is not part of a multiple system with \texttt{c1} and \texttt{c2}. However, Fig. \ref{fig:relative_distance} suggests that tidal exchanges of stars among \texttt{c1}, \texttt{c2} and \texttt{c3} may occur in the future.

\section{Summary} \label{sec:summary}

\par We have identified three OCs, named as \texttt{c1}, \texttt{c2}, \texttt{c3} in this paper. Using data from Gaia DR3, we calculate the dereddened colors $\left( G_{\mathrm{BP}}-G_{\mathrm{RP}} \right)_{0}$ and absolute magnitudes $M_{\mathrm{G}}$. Isochrone fits to the CAMD indicates that the ages of \texttt{c1}, \texttt{c2}, \texttt{c3} are about $6.3\pm0.9$, $8.1\pm1.4$ and $21.8\pm2.2$ Myr, respectively. All the cluster member candidates have similar \texttt{PMs}. From a few \texttt{RVs} in Gaia and LAMOST, we have estimated the mean \texttt{RVs} for these clusters. The difference between \texttt{c1} and \texttt{c2} is less than $3$ $\mathrm{km} \cdot \mathrm{s}^{-1}$. The Galactocentric Cylindrical velocities $\left( V_{\mathrm{R}},V_{\phi},V_{\mathrm{Z}} \right) = \left( -5.18,238.33,-7.17 \right)$ $\mathrm{km} \cdot \mathrm{s}^{-1}$, $\left( -3.54,236.09,-6.91 \right)$ $\mathrm{km} \cdot \mathrm{s}^{-1}$ and $\left( -8.42,239.61,-3.61 \right)$ $\mathrm{km} \cdot \mathrm{s}^{-1}$ for \texttt{c1}, \texttt{c2} and \texttt{c3}, respectively. The above data were used to compute the orbits of each OC and the relative distances among them. Using metallicities for individual OC stars from Gaia DR3, the median values of [M/H] are $-0.26$ dex, $-0.26$ dex, and $-0.12$ dex, respectively. From relative distances, kinematics, ages, and metallicities, we conclude \texttt{c1} and \texttt{c2} comprise a simultaneously formed primordial binary OC. We also expect there may be tidal captures among \texttt{c3} and \texttt{c1}, \texttt{c2} in the future.

\begin{deluxetable*}{cccccccccccccccccc}
  \tablenum{1}
  \tablecaption{Properties of \texttt{c1}, \texttt{c2} and \texttt{c3} from our member candidates. \label{tab:c1c2c3_properties}}
  \tablewidth{0pt}

  \tablehead{
  \colhead{Name} &
  \colhead{$gl$} &
  \colhead{$\sigma_{gl}$} &
  \colhead{$gb$} &
  \colhead{$\sigma_{gb}$} &
  \colhead{$d$} &
  \colhead{$\sigma_{d}$} &
  \colhead{$\mu_{\alpha}^{*}$} &
  \colhead{$\sigma_{\mu_{\alpha}^{*}}$} &
  \colhead{$\mu_{\delta}$} &
  \colhead{$\sigma_{\mu_{\delta}}$} &
  \colhead{\texttt{RV}} &
  \colhead{$V_{\mathrm{r}}$} &
  \colhead{$V_{\phi}$} &
  \colhead{$V_{\mathrm{Z}}$} &
  \colhead{$\tau$} & 
  \colhead{$\sigma_{\tau}$} &
  \colhead{[M/H]} \\
  \colhead{} &
  \multicolumn4c{deg} &
  \multicolumn2c{kpc} &
  \multicolumn4c{$\mathrm{mas} \cdot \mathrm{y}^{-1}$} &
  \multicolumn4c{$\mathrm{km} \cdot \mathrm{s}^{-1}$} &
  \multicolumn2c{Myr} &
  \multicolumn1c{dex}
  }
  \decimalcolnumbers
  \startdata
  \texttt{c1} & 31.28 & 0.36 & 5.25 & 0.25 & 0.426 & 0.008 & 3.24 & 0.31 & -8.59 & 0.43 & -8.05 & -5.18 & 238.33 & -7.17 & 6.3 & 0.9 & -0.26 \\
  \texttt{c2} & 31.86 & 0.49 & 2.98 & 0.37 & 0.468 & 0.017 & 2.65 & 0.27 & -8.35 & 0.27 & -9.53 & -3.54 & 236.09 & -6.91 & 8.1 & 1.4 & -0.26 \\
  \texttt{c3} & 27.17 & 0.39 & 2.59 & 0.23 & 0.377 & 0.009 & 2.06 & 0.20 & -9.00 & 0.26 & -4.15 & -8.42 & 239.61 & -3.61 & 21.8 & 2.2 & -0.12 \\
  \enddata
  \tablecomments{$\sigma_{gl}$, $\sigma_{gb}$, $\sigma_{d}$, $\sigma_{\mu_{\alpha}^{*}}$ and $\sigma_{\mu_{\delta}}$ are the standard deviations from cluster members. $\tau$ is the age of a cluster.}
\end{deluxetable*}

\acknowledgments

This study is supported by the National Natural Science Foundation of China under grant No. 11988101, 11973048, 11927804, 11890694 and National Key R\&D Program of China No. 2019YFA0405502. We acknowledge the support from the 2m Chinese Space Station Telescope project : CMS-CSST-2021-B05. This work presents results from the European Space Agency (ESA) space mission Gaia. Gaia data are being processed by the Gaia Data Processing and Analysis Consortium (DPAC). Funding for the DPAC is provided by national institutions, in particular the institutions participating in the Gaia MultiLateral Agreement (MLA). The Gaia mission website is \url{https://www.cosmos.esa.int/gaia}. The Gaia archive website is \url{https://archives.esac.esa.int/gaia}. Guoshoujing Telescope (the Large Sky Area Multi-Object Fiber Spectroscopic Telescope LAMOST) is a National Major Scientific Project built by the Chinese Academy of Sciences. Funding for the project has been provided by the National Development and Reform Commission. LAMOST is operated and managed by the National Astronomical Observatories, Chinese Academy of Sciences. TDO gratefully acknowledges support from the U.S. National Science Foundation grant AST-1910396.

\vspace{10mm}
\software{\texttt{Astropy} \citep{astropy13},
          \texttt{DBSCAN} \citep{dbscan},
          \texttt{Galpy} \citep{Bovy15},
          \texttt{HDBSCAN} \citep{Campello13,McInnes17},
          \texttt{Matplotlib} \citep{matplotlib07},
          \texttt{Numpy} \citep{numpy11},
          \texttt{Pandas} \citep{pandas10},
          \texttt{Scipy} \citep{scipy20},
          \texttt{Scikit-learn} \citep{sklearn12},
          \texttt{Topcat} \citep{topcat05}
}

\clearpage



\bibliographystyle{aasjournal}

\end{document}